\documentclass{aa}  

\usepackage{pifont}
\usepackage{amsmath}
\usepackage{graphicx}
\usepackage{txfonts}
\usepackage{url}
\usepackage{longtable}
\usepackage{lscape}
\usepackage{natbib}
\usepackage[colorlinks,linkcolor=red,anchorcolor=green,citecolor=blue]{hyperref}

\begin{document}

   \title{Unveiling Key Factors in the Solar Eruptions Leading to the Solar Superstorm in 2024 May}
  \author{Rui Wang\inst{1,2},
          Ying D. Liu\inst{1,2,3},
          Xiaowei Zhao\inst{4,5},
          \ and Huidong Hu\inst{1,2}
          }
   \institute{State Key Laboratory of Space Weather, National Space Science Center, Chinese Academy of Sciences, Beijing 100190, People's Republic of China \\ \email{rwang@swl.ac.cn} \and Solar Activity and Space Weather, National Space Science Center, Chinese Academy of Sciences, Beijing 100190, People's Republic of China \and  University of Chinese Academy of Sciences, Beijing, China \and Key Laboratory of Space Weather, National Satellite Meteorological Center (National Center for Space Weather), China Meteorological Administration, Beijing 100081, People's Republic of China \and School of Earth and Space Sciences, Peking University, Beijing 100871, People's Republic of China}

\authorrunning{R. Wang et. al.}
\titlerunning{Unveiling Key Factors in the Solar Eruptions}

   \date{Received XXX, XXX; accepted XXX, XXX}

\abstract
   {NOAA Active Region (AR) 13664/8 produced the most intense geomagnetic effects since the ``Halloween'' event of 2003. The resulting extreme solar storm is believed to be the consequence of multiple interacting coronal mass ejections (CMEs). Notably, this AR exhibites an exceptionally rapid magnetic flux emergence. The eruptions we are focusing on all occurred along collisional polarity inversion lines (PILs) through ``collisional shearing'' during a three-day period of extraordinarily high flux emergence ($\sim$10$^{21}$ Mx hr$^{-1}$). Our key findings reveal how photospheric magnetic configurations in eruption sources influence solar superstorm formation and geomagnetic responses, and link exceptionally strong flux emergence to sequential homologous eruptions: (1) We identified the source regions of seven halo CMEs, distributed primarily along two distinct PILs, suggesting the presence of two groups of homologous CMEs. (2) The variations in magnetic flux emergence rates at the source regions correlate with CME intensities, potentially explaining the two contrasting cases of complex ejecta observed at Earth. (3) Calculations of magnetic field gradients around CME source regions show strong correlations with eruptions, providing crucial insights into solar eruption mechanisms and enhancing future prediction capabilities.}


   \maketitle
%

\section{Introduction}
The impact of space weather on human society is increasingly significant. The influence of solar superstorms stands out due to its remarkable geomagnetic effects. \citet{2014Liuxying2,2019Liuxying} point out that solar superstorms often form through a mechanism known as a ``perfect storm'', where multiple factors combine to amplify the intensity of a storm that would otherwise be moderate. These factors typically include the consecutive eruption of multiple coronal mass ejections (CMEs), preconditioning, and CME-CME interactions. \citet{2019Liuxying} emphasize the particular importance of preconditioning in generating Carrington-class solar storms.

These consecutive CMEs often originate from the same active region (AR) or even the same polarity inversion line (PIL), commonly known as homologous CMEs. They often form in new or highly dynamic ARs exhibiting strong magnetic flux emergence. To initiate a solar eruption, two essential factors are typically required: the presence of a magnetic flux-rope (MFR) structure in the eruption source region and the triggering of this MFR by instability. MFRs can form through two primary mechanisms. Strong photospheric shearing motions, through the tether-cutting process, can create a twisted MFR \citep{1989Vanballegooijen,2001Moore}. In this process, magnetic reconnection occurs, causing a significant amount of magnetic flux to be cancelled and submerged below the photosphere, and the coronal magnetic field around the PIL becomes more sheared and twisted. On the other hand, evidence suggests that twisted magnetic structures already exist beneath the photosphere, and during the process of flux emergence, these twisted structures directly rise to the photosphere as MFRs \citep{2008Okamoto}. Once the MFR has formed, instability becomes the key factor leading to the eventual eruption. It can arise from the MFR itself, such as when a twisted MFR exceeds a critical value of twist, triggering kink instability \citep{2004Fan,2004Kliem,2004Torok}. It also arises from the confining magnetic fields around the MFR, such as torus instability. This occurs when the Sun-directed Lorentz force decreases faster with height than the radial outward-directed ``hoop force'' \citep{2006Kliem}.

\citet{2019Chintzoglou} propose a scenario called ``collisional shearing''. This scenario suggests that, accompanying the dynamic flux emergence, collisional-shearing processes occur between nonconjugated magnetic polarities. Nonconjugated polarities refer to the interactions between two non-dipolar magnetic polarity pairs with opposite polarities. According to \citet{2019Chintzoglou}, they indicate that most of the consecutive eruptions occur at these collisional-PILs rather than at the PILs of dipolar pairs. Their approach allows for a more accurate measurement of the cancelled magnetic flux during flux emergence. \citet{2022Rui} apply this method to measure the cancelled magnetic flux in AR 11283 and estimate that the cancelled flux accounted for over 24\% of the total unsigned flux. This subtantial cancelled flux explains why this AR is able to produce four consecutive major eruptions within a short span of three days. \citet{2023Dhakal} utilize the method proposed by \citet{2007Schrijver} to extract a strong gradient PIL (SgPIL) and found a strong correlation between SgPIL and flare productivity for super-active ARs. Indeed, the formation of the SgPIL is strongly associated with the collisional-shearing scenario.

The series of eruption events that occurred in May 2024 resulted in a strong geomagnetic response close to that of the ``Halloween'' event in 2003, with the Dst index reaching $\sim$-412 nT. These eruptive events are associated with AR 13664/8, which exhibited exceptionally strong magnetic flux emergence, producing multiple M-class and above flares, each accompanied by halo CMEs. The intense flux emergence led to an unusually complex magnetic field configuration in the region. In this letter, we mainly focus on how photospheric magnetic configurations in eruption sources influence solar superstorm formation and geomagnetic responses, and link exceptionally strong flux emergence to sequential homologous eruptions. In Section 2, we present the data analysis of the magnetic field and imaging results, and in Section 3, we draw conclusions and engage in discussions.

\section{Data Analysis and Results}\label{sec2}
Solar storms have caused prolonged geomagnetic disturbances, with the Dst index rapidly decreasing to $\sim$-412 nT on 2024 May 10, and failing to fully recover to normal levels even after May 13. In their recent study, \citet{2024Liuxying} analyze the data and identify a series of consecutive halo CMEs, some of which exhibite a strong southward component. They categorize these halo CMEs into two contrasting cases of complex ejecta. It is determined that the four consecutive halo CMEs from May 8 to 9 are the primary contributors to the rapid Dst decline on May 10 (ref. GOES soft X-ray cuvrve of Figure 1). Subsequently, three halo eruptions between May 9 17:44 UT and May 11 were identified as the main factors responsible for the sustained and incomplete recovery of Dst. Readers are directed to \citet{2024Liuxying} for the indentification and discussions of the full halo CMEs. The characteristics of the source region for these eruptions are yet to be determined.

\begin{figure*}[h]
    \centering
    \includegraphics[width=1.0\textwidth]{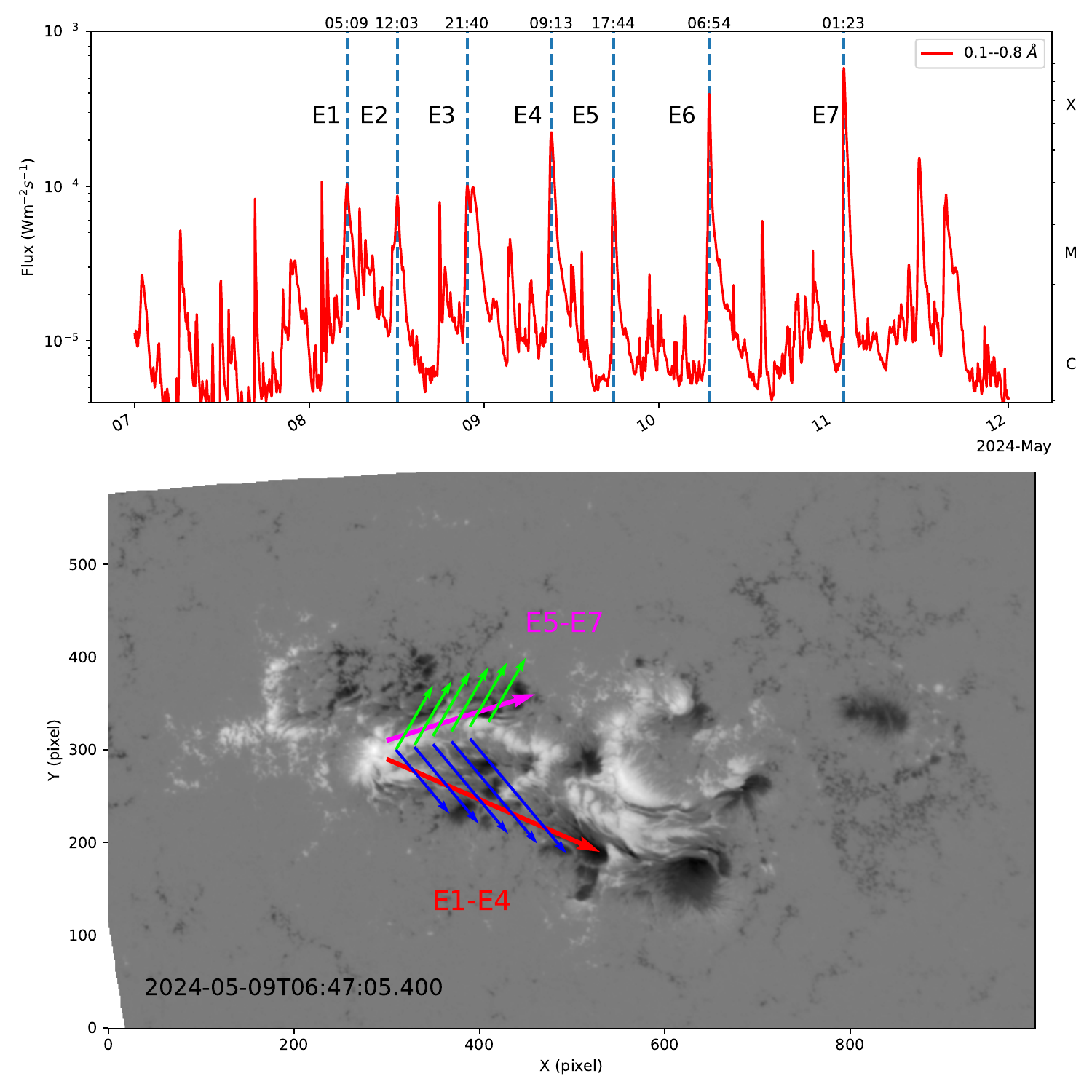}
    \caption{GOES 1-8 \AA~flux (top) and AR 13664/8 overview (bottom). Vertical blue dashed lines: flare peak times. Arrows: axial fields of potential MFR structures along different PILs (red/purple), with reverse azimuthal fields (blue/green).}\label{fig1}
\end{figure*}

Through careful examination, we have determined that the first four eruptions originate from the lower region indicated in Figure 1, with the axial magnetic field of the MFR aligned roughly along the red arrow and the azimuthal magnetic field along the blue arrows. The southward field distribution explains why these initial four eruptions led to a rapid decrease in the Dst index. The subsequent three eruptions originate from the upper region, with the axial field aligned in the direction of the purple arrow, and the azimuthal field also undergoes a significant reversal, approaching a northward magnetic field orientation. The subsequent eruptions which continue to include X-class eruptions after May 11 are not further discussed due to their proximity to the limb or on the far side of the Sun.

\begin{figure*}[h]
    \centering
    \includegraphics[width=1.0\textwidth]{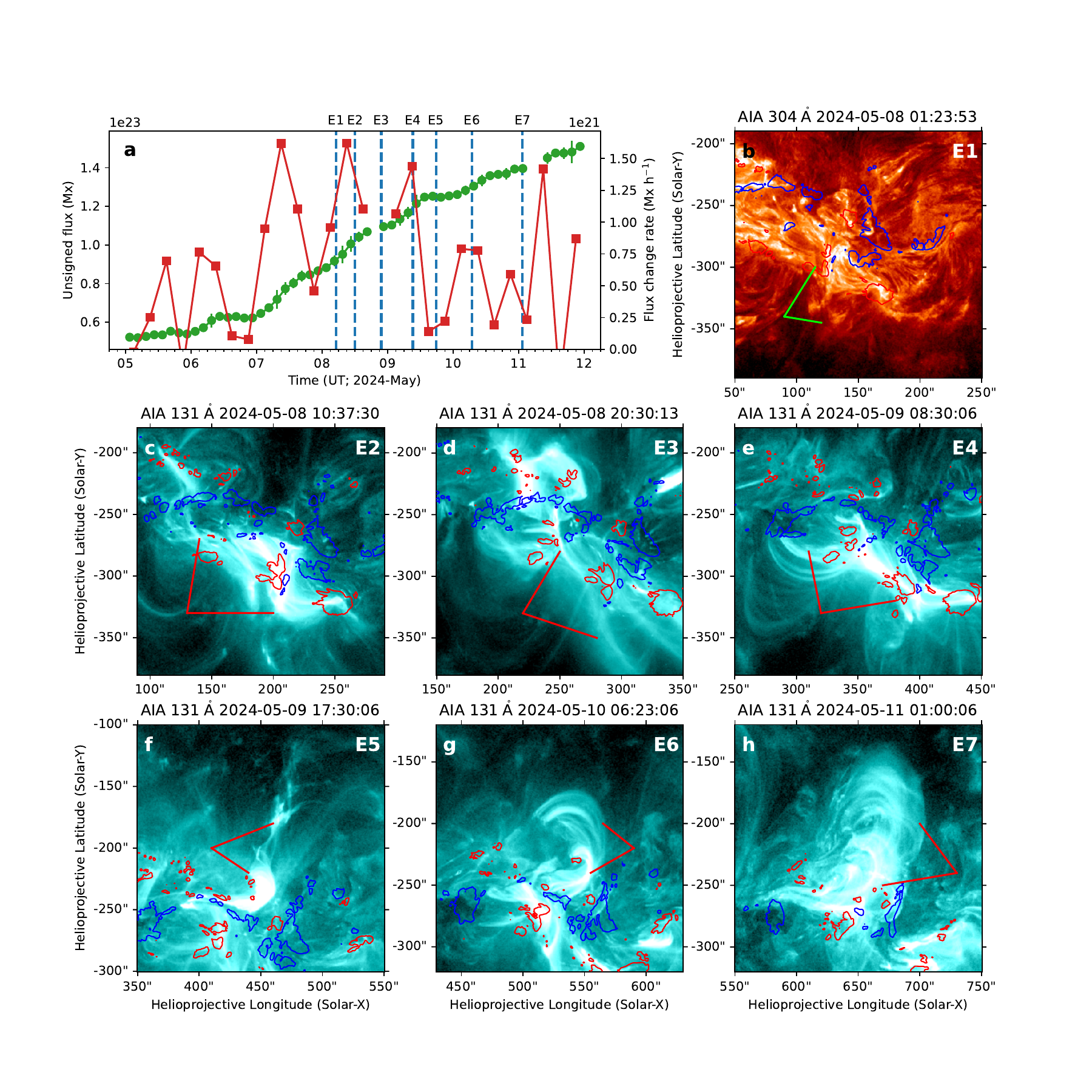}
    \caption{(a) Mean unsigned magnetic flux (green, 3-hour windows) and 6-hour average flux emergence rate of AR 13664/8 (red). Error bars: 3$\sigma$ within each window. (b)-(h) The CME (E1-E7) source regions in AIA 304 \AA~and 131 \AA. The green and red bars point to filament/hot channel locations. Positive (blue) and negative (red) fluxes ($\pm$1000 G) are overplotted.}\label{fig2}
\end{figure*}

This AR is formed by the convergence of two neighboring ARs (AR 13664/8), through a process known as ``collisional shearing''. The complex AR is undergoing fast flux emergence, with a continuous increase in unsigned flux and exhibiting an extraordinary magnetic flux emergence rate (see Figure 2a). We calculate the unsigned magnetic flux for May 5-11 from the Helioseismic and Magnetic Imager \citep[HMI;][]{2012Scherrer,2012Schou} data series ``hmi.sharp\_cea\_720s\_dconS'', which offers improved quality for flux calculations. Our analysis focuses on pixels with field strengths exceeding 200 G. The unsigned flux reached a maximum of $\sim$1.5$\times$10$^{23}$ Mx by the end of May 11. Most remarkably, the 6-hour average flux emergence rate peaks at an exceptional $\sim$1.6$\times$10$^{21}$ Mx hr$^{-1}$ on May 8. This rate is truly extraordinary, surpassing even the most notable ARs of recent solar cycles, including AR 12673 -- responsible for the most intense flare of cycle 24, and AR 12192 -- the largest sunspot group observed since 1990. In the Stanford HMI science nuggets, Sun et al (2024) \footnote{\url{http://hmi.stanford.edu/hminuggets/?p=4216}} indicate that this event likely represents the most rapid flux emergence ever recorded in the SDO era, marking a historic milestone in solar observations. Our findings, while showing slightly lower peak unsigned magnetic flux, strongly support this assertion. The emergence rate profile reveals a primary peak on May 8, flanked by substantial secondary peaks on May 7 and 9 (Figure 2a). Notably, the average emergence rate over this three-day period sustained an extraordinarily high magnitude of $\sim$10$^{21}$ Mx hr$^{-1}$.

When adjusting the magnetic field threshold to 150 G, our results align closely with those reported by Sun et al. (2024), further confirming the historical significance of this event. This extraordinary flux emergence rate not only sets a new benchmark for solar activity observations but also challenges our understanding of the underlying mechanisms driving such extreme events.

\begin{figure*}[h]
    \centering
    \includegraphics[width=1.0\textwidth]{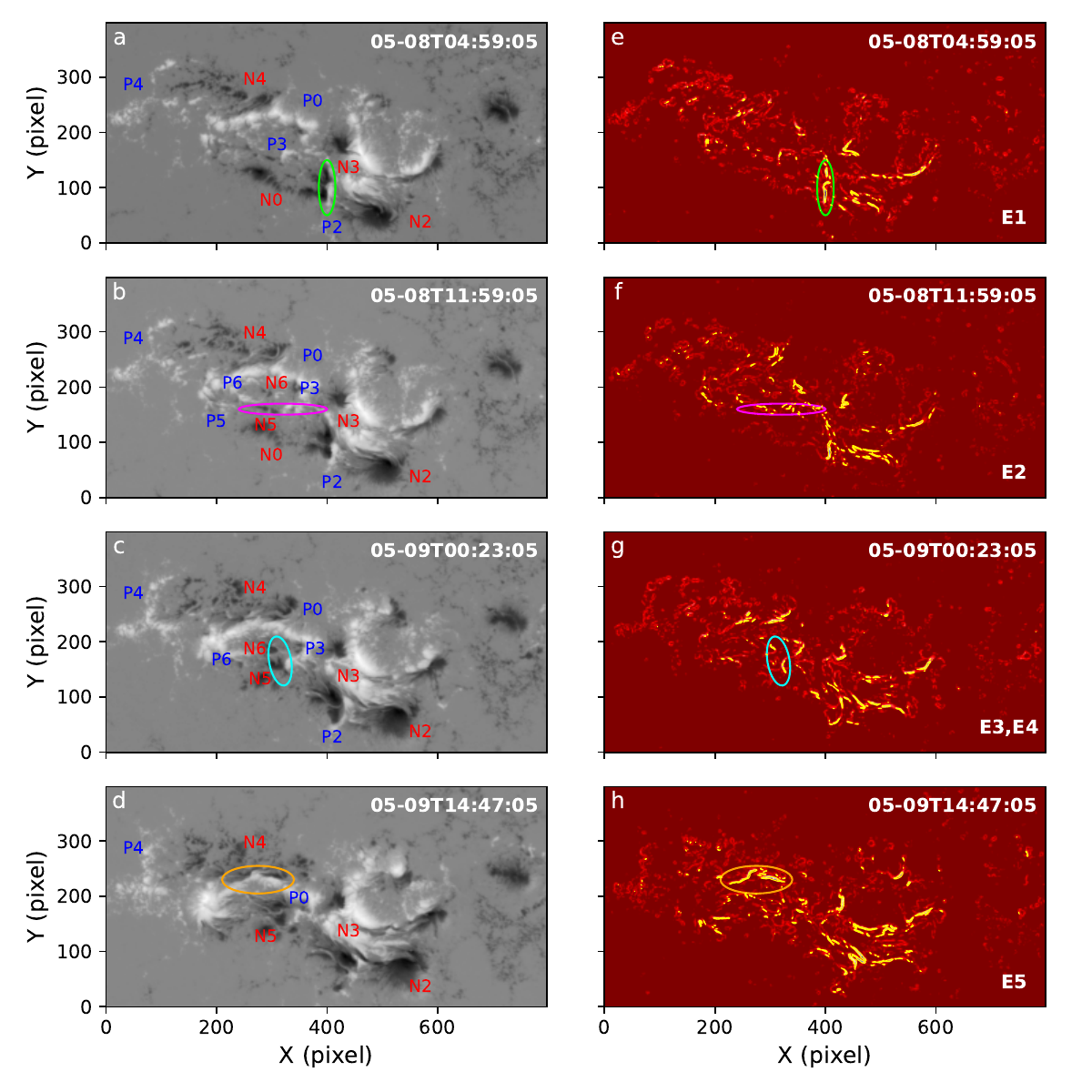}
    \caption{B$_r$ magnetograms (a-d) and their corresponding gradient maps (e-h) at sequential times. Color eclipses indicates the CME source regions. Dipolar magnetic fields are shown. Yellow contours in (e-h) represents strong gradient ($\geq$ 1000 G Mm$^{-1}$). Left and right columns correspond to two separate animations. A five-day animation of the magnetograms and gradient maps is available as an online movie, starting at 2024 May 4 23:59 UT, and ending at 2024 May 9 23:47 UT.}\label{fig3}
\end{figure*}

We focus on five eruptions (E1-E5) from May 8 to 9 within $\pm$30$^\circ$ longitude, where magnetic field data is considered more reliable \citep{2014Hoeksema}, while providing source imaging for two additional eruptions (E6, E7) outside this range as reference. These correspond to CMEs 1-7 in \citet{2024Liuxying}. From the Extreme Ultraviolet (EUV) imaging data captured by the Atmospheric Imaging Assembly \citep[AIA;][]{2012Lemen}, we observe a filament or hot channel structures in the eruption source region (Figure 2b-2h). Except for E1, where the filament is visible in the relatively cooler 304 \AA~wavelength channel (along the north-south direction), the remaining eruptions can only be observed in the hotter 131 \AA, indicating that the continuous eruptions have heated up the source region significantly. By analyzing the onset of EUV brightening, we can roughly determine the location of the PILs corresponding to the eruption source region. E5 exhibits a large-scale hot channel structure, but the primary eruption source region is likely associated with the brightening structure above the strong magnetic field region.

Figure 3 presents the magnetograms and magnetic field gradient distribution corresponding to the eruptions. We utilized the line-of-sight (LOS) magnetic field data instead of Spaceweather HMI Active Region Patches \citep[SHARPs;][]{2014Bobra,2014Hoeksema} for two reasons: firstly, the scientific SHARP data is currently unavailable, and secondly, the SHARP data for May 8 has several hours of missing observations. The LOS magnetic field data provides a continuous and reliable alternative for our analysis. We follow the methodology used for processing SHARP data to handle the LOS magnetic field data. Initially, we preprocess full-disk 45 s LOS magnetograms at a time interval of 720 s, to account for satellite rotation that occurred between 16:00 and 24:00 UT on May 8. Subsequently, we create cutout maps from these full-disk LOS observations, with a field of view of 1000$\times$600 pixels, large enough to encompass the AR, while co-moving with the guiding center of the AR. To transform the cutout maps from the native helioprojective coordinate system to a local Cartesian coordinate system, we remapped the data using a cylindrical equal area (CEA) projection. Additionally, it is important to note that the observed LOS field is not truly radial but rather projected onto the LOS. To obtain the radial magnetic field values B$_r$, we divided the original data B$_{los}$ by $\mu$, where $\mu$ represents the cosine of the angle between the LOS and the local normal at the solar surface.

\begin{figure*}[h]
    \centering
    \includegraphics[width=1.0\textwidth]{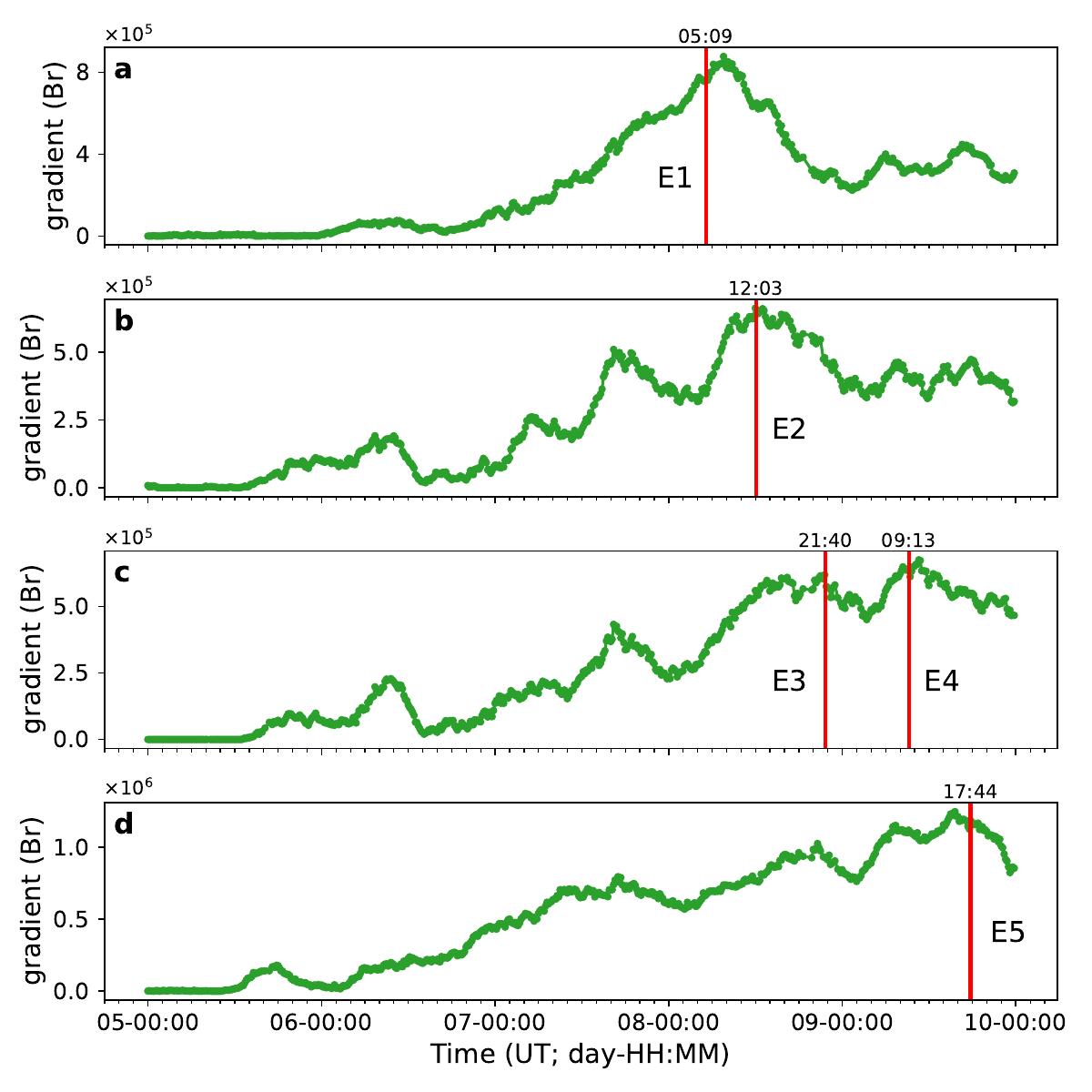}
    \caption{Cumulative sum of strong gradients within the elliptical regions shown in Figure 3. Red vertical lines correpond to E1-E5.}\label{fig4}
\end{figure*}

The left column of Figure 3 displays magnetograms, with ellipses of different colors outlining the eruption source regions. These regions correspond to the nonconjugated collisional-PILs of two sets of emerging dipolar magnetic fields. The elliptise is designed to maximally encompass the resultant high-gradient PIL, enabling full tracking of magnetic field gradient variations by collisional shearing. The right column shows the B$_r$ gradient map with a gradient exceeding the threshold of 500 G Mm$^{-1}$. Additionally, regions with a gradient exceeding 1000 G Mm$^{-1}$ are marked with yellow contours, representing strong gradient areas. By comparing the maps in the left and right columns, it is evident that almost all eruption source regions exhibit high gradients along the collisional-PILs. We calculate the cumulative sum of strong gradients within the elliptical regions. Figure 4 demonstrates that each eruption (vertical lines indicating peak times of the flares) is accompanied by an increase in the cumulative gradient, which then gradually decreases after the eruption. This illustrates a strong correlation between magnetic gradients and solar eruptions.

The AR exhibites a historically high magnetic flux emergence rate, with a maximum rate of approximately 1.6$\times$10$^{21}$ Mx hr$^{-1}$ over 6-hour intervals. It is worth noting that the magnetic field strength of this AR remains below 2000 Gauss, which is significantly lower compared to ARs with stronger magnetic field strength \citep[e.g., AR 11944, $>$ 3000 G;][]{2015Rui}. This could possibly be attributed to the exceptionally active flux emergence, which impedes the process of magnetic field concentration into high-strength polarities locally.

AR 13664 is pre-existing, with its positive and negative magnetic polarites moving towards the southeast-northwest direction on both sides. Subsequently, AR 13668 emerges to the east of AR 13664, undergoing a complex merging process with AR 13664, resulting in the formation of a more complex and large AR. We distinguish between dipole and non-dipole field collisions through visual tracking of magnetic poles. When positive and negative poles appear simultaneously and move synchronously in opposite directions, we identify them as a dipole pair. Since our goal is merely to identify dipole pairs rather than perform magnetic field calculations (e.g., magnetic flux cancellation), visual tracking suffices for this purpose. The early interaction is shown in Figure 2b at locations (120$^{\prime\prime}$, -300$^{\prime\prime}$). E1 occurred at the nonconjugated magnetic PILs between P2-N3/N0 (see Figure 3a) oriented in the north-south direction. Subsequently, E2 occurred at locations (170$^{\prime\prime}$, -260$^{\prime\prime}$; Figure 2c). The B$_r$ components of the two sets of nonconjugated magnetic polarities (P6-N5; Figure 3b) are not very strong. Sun et al. (2024) reveal that the penumbra area of this region gradually expands, implying that the magnetic field becomes more horizontal, with an increasing inclination angle. Such a distribution of magnetic fields on the photosphere is favorable for the formation of MFRs.

As mentioned before, the generation of large eruptions requires both the buildup of nonpotentiality and triggering factors. The collision of nonconjugated magnetic fields (P3-N5) may serve as a trigger factor, as a flare E2 occurred directly above their convergence region while P3 and N5 continuously converged. It is likely associated with tether-cutting reconnection of reverse coronal magnetic fields above the PIL of P3-N5. We use a horizontal eclipse to calculate the total gradient of E2 source region in Figure 3. The left side of the ellipse exhibits evident shearing motion of P6 and N5 (Figure 3b), providing the necessary conditions for the buildup of MFRs \citep{2019Chintzoglou,2022Rui}. The occurrences of E3 and E4 can be seen as a sustained progression of the collisional-shearing process that gave rise to E2. Once again, this pair of nonconjugated magnetic polarities (P3-N5/N6) interacted at positions (250$^{\prime\prime}$, -270$^{\prime\prime}$) and (350$^{\prime\prime}$, -270$^{\prime\prime}$), resulting in E3 and E4 (see Figure 2d and 2e). In fact, E2-E4 can be considered as a series of homologous eruptions. Notably, Figure 2a reveals an exceptionally high average magnetic flux emergence rate of $\sim$10$^{21}$ Mx hr$^{-1}$ during May 7-9. The peak emergence rate occurred on May 8 just before E2, then followed by the other two consecutive homologous eruptions. 

As for E5, its location (410$^{\prime\prime}$, -240$^{\prime\prime}$) shiftes to a more northern PIL (see Figure 2f). The lower panel of Figure 1 indicates that the azimuthal field around this set of PILs has a reverse direction with the lower set. From Figures 3a-d, it shows that a pair of conjugated magnetic polarities (P4-N4) undergoes shearing motion, with the negative polarity N4 moving westward relative to the positive polarity. As the magnetic field gradient continues to rise (Figure 4d), E5 was triggered when a negative polarity rapidly intrude into the positive polarity P0. E6 and E7 share the same PIL as E5. This PIL subsequently produced two even stronger X-class flares, X3.9 and X5.8 (as shown in Figures 2g and 2h). Figure 2a indicates that these two eruptions coincide with periods of continued high magnetic flux emergence rates. However, as these two eruptions occur in ARs beyond 30$^\circ$W longitude, where magnetic field data become unreliable, we omit gradient calculations for them.

\section{Conclusion and Discussion} 
We have examined the solar sources of the 2024 May geomagnetic superstorm, the largest in two decades. Our key findings reveal how photospheric magnetic configurations in eruption sources influence solar superstorm formation and geomagnetic responses, and link exceptionally strong flux emergence to sequential homologous eruptions. Our results are summarized as follows:

(1) Seven eruptions (E1-E7) occurred mainly along two primary PILs, as shown in Figure 1. Comparing the AIA 131 \AA~high-temperature channel observations (Figure 2), we believe these source regions likely contain MFRs or sheared arcades as the initial magnetic field configuration for CMEs. Based on the photospheric magnetic field distribution, we can roughly divide the source regions into two groups of homologous CMEs. The first group (E1-E4) has axial field generally aligned with the red arrow and poloidal field along the blue arrows, both dominated by southward components. The second group (E5-E6) shows predominantly northward components in both axial and poloidal fields. \citet{2024Liuxying} identified two contrasting cases of complex ejecta in terms of their geo-effectiveness from in situ data. The first complex ejecta exhibits strong southward magnetic field components, contributing to the rapid Dst index decrease on May 10. The second shows weak southward fields, resulting in a slow Dst index recovery after May 13. The different magnetic field distributions around the eruption sources may explain their distinct geo-effectiveness, despite both resulting from CME-CME interactions.

(2) Active region 13664/8 experienced historically strong magnetic flux emergence, peaking on May 8 at an exceptional rate of $\sim$1.6$\times$10$^{21}$ Mx hr$^{-1}$. From May 7-9, the average emergence rate sustained an extraordinarily high magnitude of $\sim$10$^{21}$ Mx hr$^{-1}$. \citet{2024Liuxying} note that the CME associated with E5 exhibits a marked reduction in velocity relative to its predecessor. This velocity difference is believed to be directly linked to the formation of the two distinctly different complex ejecta mentioned above. According to CME-CME interaction theory, the trailing CME needs to be faster than the leading one to catch up and interact. Figure 2a shows that E5 (X1.1 flare, 940 km/s, see Table 1 of \citealt{2024Liuxying}) occurred after the three-day period of strongest average emergence rate, with both the associated flare and CME speed lower than E4 (X2.2 flare, 1480 km/s). The significant drop in emergence rate before E5 may explain its slower CME speed. The subsequent increase in emergence rate for the following two eruptions demonstrates a correlation between eruption intensity and flux emergence rate.

(3) Detailed analysis of photospheric magnetic field evolution near the erupting PILs reveals that these PILs can be classified as collisional-PILs proposed by \citet{2019Chintzoglou}. Our results indicate strong magnetic field gradients along these collisional-PILs, suggesting a connection between gradient enhancement and eruption occurrence. The magnetic field gradients surrounding the PILs for all eruptions peak at the onset of each event, followed by a subsequent decrease in intensity. The gradient enhancement at PILs is related to the converging motion of opposite polarity fields on both sides of the PIL, driven by the collisional-shearing process, which is induced by the super-strong magnetic flux emergence. This clarifies the underlying cause of the correlation between eruption intensity and emergence rate mentioned in (2). The collisional shearing efficiently converts magnetic flux emergence into rapid accumulation of nonpotentiality necessary for eruptions. The continuous increase in magnetic field gradients reflects the cumulative effect of emerging flux at the PILs.

The choice of magnetic field gradients as a key factor in assessing the eruptions is motivated by the direct consequences of the collisional-shearing mechanism, which lead to the compression and increased densities of the magnetic fields on both sides of the nonconjugated magnetic polarities. Althought magnetic shear angle typically increases concomitantly, its variations do not always correspond directly with changes in the continuously emerging magnetic flux. When opposite magnetic poles are in close proximity, the shear angle tends to stabilize. Consequently, quantifying the impact of rapid magnetic flux emergence on eruptions is more effectively achieved by analyzing the changes in magnetic field gradients around PILs. We believe the continuous increase in gradient prior to eruption is closely linked to the buildup of MFRs at the collisional-PILs. When this gradient reaches a certain threshold, further convergence of the two groups of magnetic field likely induces local reconnection, initiating the eruption. However, collisional shearing does not always simultaneously contribute to both non-potential field buildup and eruption triggering. In most cases, they mainly affect the accumulation of nonpotentiality. Triggering factors are more complex, including previously mentioned kink and torus instabilities.

Factor such as ``MEANGBH'', identified by \citet{2016Bobra} as a significant influence on CME prediction, as well as earlier studies by \citet{2007Schrijver} on the R factor and \citet{2023Dhakal} on SgPILs, all focus on changes in the magnetic field gradients along the PIL resulting from active flux emergence. These factors, along with their derived quantities are often used to evaluate triggering mechanisms in ARs \citep[e.g.,][]{2012Petrie,2015SunXD,2017Vemareddy,2018Rui2,2022Rui,2022Ran}. In practice, there is often a focus on overall gradient parameters of the AR or all the PILs in general, without specifically identifying which high-gradient PILs are associated with eruptions. As shown in Figures 3e-h, apart from the eruption source region, high-gradient PILs are actually observed at various locations. If we can accurately measure the gradient changes directly associated with the source region, we believe it will demonstrate a stronger correlation with solar eruptions. Additionally, to achieve a higher correlation, it is crucial to analyze large-scale eruptions that display substantial space weather effects and exhibit homologous properties in their eruption source regions. In this way, the stronger correlation can be better reflected. However, this may not facilitate prediction, as we are not entirely certain which locations will experience eruptions. A better approach would be to focus on regions where noticeable collisional-shearing processes occur. If such a region has already experienced significant eruptions, there is a high likelihood of subsequent eruptions. We will consider conducting research in this area in the future.

\begin{acknowledgements}
This research was supported by the Strategic Priority Research Program of the Chinese Academy of Sciences No.XDB0560000, NSFC under grant 12073032, the National Key R\&D Program of China No. 2022YFF0503800 and No. 2021YFA0718600, and the Specialized Research Fund for State Key Laboratories of China. Y.D.L., X.W.Z., and H.D.H. also acknowledge support from NSFC under grants 42274201, 42204176, and 42150105, respectively. We acknowledge the use of data from SDO.
\end{acknowledgements}

\bibliographystyle{aa}
\bibliography{references}

\begin{thebibliography}{26}
\expandafter\ifx\csname natexlab\endcsname\relax\def\natexlab#1{#1}\fi

\bibitem[{{Bobra} \& {Ilonidis}(2016)}]{2016Bobra}
{Bobra}, M.~G. \& {Ilonidis}, S. 2016, \apj, 821, 127

\bibitem[{{Bobra} {et~al.}(2014){Bobra}, {Sun}, {Hoeksema}, {Turmon}, {Liu},
  {Hayashi}, {Barnes}, \& {Leka}}]{2014Bobra}
{Bobra}, M.~G., {Sun}, X., {Hoeksema}, J.~T., {et~al.} 2014, \solphys, 289,
  3549

\bibitem[{{Chintzoglou} {et~al.}(2019){Chintzoglou}, {Zhang}, {Cheung}, \&
  {Kazachenko}}]{2019Chintzoglou}
{Chintzoglou}, G., {Zhang}, J., {Cheung}, M. C.~M., \& {Kazachenko}, M. 2019,
  \apj, 871, 67

\bibitem[{Dhakal \& Zhang(2023)}]{2023Dhakal}
Dhakal, S.~K. \& Zhang, J. 2023, The Astrophysical Journal, 960, 36

\bibitem[{{Fan} \& {Gibson}(2004)}]{2004Fan}
{Fan}, Y. \& {Gibson}, S.~E. 2004, \apj, 609, 1123

\bibitem[{{Hoeksema} {et~al.}(2014){Hoeksema}, {Liu}, {Hayashi}, {Sun},
  {Schou}, {Couvidat}, {Norton}, {Bobra}, {Centeno}, {Leka}, {Barnes}, \&
  {Turmon}}]{2014Hoeksema}
{Hoeksema}, J.~T., {Liu}, Y., {Hayashi}, K., {et~al.} 2014, \solphys, 289, 3483

\bibitem[{Kliem {et~al.}(2004)Kliem, Titov, \& T{\"o}r{\"o}k}]{2004Kliem}
Kliem, B., Titov, V., \& T{\"o}r{\"o}k, T. 2004, Astronomy \& Astrophysics,
  413, L23

\bibitem[{{Kliem} \& {T{\"o}r{\"o}k}(2006)}]{2006Kliem}
{Kliem}, B. \& {T{\"o}r{\"o}k}, T. 2006, Physical Review Letters, 96, 255002

\bibitem[{{Lemen} {et~al.}(2012){Lemen}, {Title}, {Akin}, {Boerner}, {Chou},
  {Drake}, {Duncan}, {Edwards}, {Friedlaender}, {Heyman}, {Hurlburt}, {Katz},
  {Kushner}, {Levay}, {Lindgren}, {Mathur}, {McFeaters}, {Mitchell}, {Rehse},
  {Schrijver}, {Springer}, {Stern}, {Tarbell}, {Wuelser}, {Wolfson}, {Yanari},
  {Bookbinder}, {Cheimets}, {Caldwell}, {Deluca}, {Gates}, {Golub}, {Park},
  {Podgorski}, {Bush}, {Scherrer}, {Gummin}, {Smith}, {Auker}, {Jerram},
  {Pool}, {Soufli}, {Windt}, {Beardsley}, {Clapp}, {Lang}, \&
  {Waltham}}]{2012Lemen}
{Lemen}, J.~R., {Title}, A.~M., {Akin}, D.~J., {et~al.} 2012, \solphys, 275, 17

\bibitem[{{Liu} {et~al.}(2014){Liu}, {Luhmann}, {Kajdi{\v{c}}}, {Kilpua},
  {Lugaz}, {Nitta}, {M{\"o}stl}, {Lavraud}, {Bale}, {Farrugia}, \&
  {Galvin}}]{2014Liuxying2}
{Liu}, Y.~D., {Luhmann}, J.~G., {Kajdi{\v{c}}}, P., {et~al.} 2014, Nature
  Communications, 5, 3481

\bibitem[{{Liu} {et~al.}(2019){Liu}, {Zhao}, {Hu}, {Vourlidas}, \&
  {Zhu}}]{2019Liuxying}
{Liu}, Y.~D., {Zhao}, X., {Hu}, H., {Vourlidas}, A., \& {Zhu}, B. 2019, \apjs,
  241, 15

\bibitem[{{Liu et al.}(2024)}]{2024Liuxying}
{Liu et al.}, Y.~D. 2024, ApJL, submitted

\bibitem[{{Moore} {et~al.}(2001){Moore}, {Sterling}, {Hudson}, \&
  {Lemen}}]{2001Moore}
{Moore}, R.~L., {Sterling}, A.~C., {Hudson}, H.~S., \& {Lemen}, J.~R. 2001,
  \apj, 552, 833

\bibitem[{{Okamoto} {et~al.}(2008){Okamoto}, {Tsuneta}, {Lites}, {Kubo},
  {Yokoyama}, {Berger}, {Ichimoto}, {Katsukawa}, {Nagata}, {Shibata},
  {Shimizu}, {Shine}, {Suematsu}, {Tarbell}, \& {Title}}]{2008Okamoto}
{Okamoto}, T.~J., {Tsuneta}, S., {Lites}, B.~W., {et~al.} 2008, \apjl, 673,
  L215

\bibitem[{{Petrie}(2012)}]{2012Petrie}
{Petrie}, G.~J.~D. 2012, \apj, 759, 50

\bibitem[{Ran {et~al.}(2022)Ran, Liu, Guo, \& Wang}]{2022Ran}
Ran, H., Liu, Y.~D., Guo, Y., \& Wang, R. 2022, The Astrophysical Journal, 937,
  43

\bibitem[{{Scherrer} {et~al.}(2012){Scherrer}, {Schou}, {Bush}, {Kosovichev},
  {Bogart}, {Hoeksema}, {Liu}, {Duvall}, {Zhao}, {Title}, {Schrijver},
  {Tarbell}, \& {Tomczyk}}]{2012Scherrer}
{Scherrer}, P.~H., {Schou}, J., {Bush}, R.~I., {et~al.} 2012, \solphys, 275,
  207

\bibitem[{{Schou} {et~al.}(2012){Schou}, {Scherrer}, {Bush}, {Wachter},
  {Couvidat}, {Rabello-Soares}, {Bogart}, {Hoeksema}, {Liu}, {Duvall}, {Akin},
  {Allard}, {Miles}, {Rairden}, {Shine}, {Tarbell}, {Title}, {Wolfson},
  {Elmore}, {Norton}, \& {Tomczyk}}]{2012Schou}
{Schou}, J., {Scherrer}, P.~H., {Bush}, R.~I., {et~al.} 2012, \solphys, 275,
  229

\bibitem[{{Schrijver}(2007)}]{2007Schrijver}
{Schrijver}, C.~J. 2007, \apjl, 655, L117

\bibitem[{{Sun} {et~al.}(2015){Sun}, {Bobra}, {Hoeksema}, {Liu}, {Li}, {Shen},
  {Couvidat}, {Norton}, \& {Fisher}}]{2015SunXD}
{Sun}, X., {Bobra}, M.~G., {Hoeksema}, J.~T., {et~al.} 2015, \apjl, 804, L28

\bibitem[{T{\"o}r{\"o}k {et~al.}(2004)T{\"o}r{\"o}k, Kliem, \&
  Titov}]{2004Torok}
T{\"o}r{\"o}k, T., Kliem, B., \& Titov, V. 2004, Astronomy \& Astrophysics,
  413, L27

\bibitem[{{van Ballegooijen} \& {Martens}(1989)}]{1989Vanballegooijen}
{van Ballegooijen}, A.~A. \& {Martens}, P.~C.~H. 1989, \apj, 343, 971

\bibitem[{Vemareddy(2017)}]{2017Vemareddy}
Vemareddy, P. 2017, The Astrophysical Journal, 851, 3

\bibitem[{Wang {et~al.}(2015)Wang, Liu, Dai, Yang, Huang, \& Hu}]{2015Rui}
Wang, R., Liu, Y.~D., Dai, X., {et~al.} 2015, The Astrophysical Journal, 814,
  80

\bibitem[{{Wang} {et~al.}(2018){Wang}, {Liu}, {Hoeksema}, {Zimovets}, \&
  {Liu}}]{2018Rui2}
{Wang}, R., {Liu}, Y.~D., {Hoeksema}, J.~T., {Zimovets}, I.~V., \& {Liu}, Y.
  2018, \apj, 869, 90

\bibitem[{{Wang} {et~al.}(2022){Wang}, {Liu}, {Yang}, \& {Hu}}]{2022Rui}
{Wang}, R., {Liu}, Y.~D., {Yang}, S., \& {Hu}, H. 2022, \apj, 925, 202

\end{thebibliography}

\end{document}